\documentclass[12pt]{article}
\usepackage{amsmath,amssymb,graphicx,mathrsfs,hyperref,slashed}
\newcommand{\be}{\begin{equation}}
\newcommand{\ee}{\end{equation}}
\newcommand{\bea}{\begin{eqnarray}}
\newcommand{\eea}{\end{eqnarray}}

\newcommand{\wh}{\widehat}
\newcommand{\wt}{\widetilde}

\newcommand{\ww} {-\omega}
\def\({\left(} \def\){\right)}
\def\N{{\cal N}}
\renewcommand{\baselinestretch}{1.15}
\begin{document}

\title{\vspace{-1.8in}
\begin{flushright} {\footnotesize LMU-ASC 72/13 }  \end{flushright}
\vspace{3mm}
\vspace{0.3cm} {Horizons of semiclassical black holes are cold}}
\author{\large Ram Brustein${}^{(1,2)}$,  A.J.M. Medved${}^{(3)}$ \\
\vspace{-.5in} \hspace{-1.5in} \vbox{
 \begin{flushleft}
  $^{\textrm{\normalsize
(1)\ Department of Physics, Ben-Gurion University,
    Beer-Sheva 84105, Israel}}$
  $^{\textrm{\normalsize  (2) CAS, Ludwig-Maximilians-Universit\"at M\"unchen, 80333 M\"unchen, Germany}}$
$^{\textrm{\normalsize (3)  Department of Physics \& Electronics, Rhodes University,
  Grahamstown 6140, South Africa }}$
 \\ \small \hspace{1.7in}
    ramyb@bgu.ac.il,\  j.medved@ru.ac.za
\end{flushleft}
}}
\date{}
\maketitle

 \begin{abstract}

We  calculate, using our recently proposed semiclassical framework, the quantum state of the Hawking pairs that are produced during the evaporation of a black hole (BH). Our framework adheres to the standard rules of quantum mechanics and incorporates the quantum fluctuations of the collapsing shell spacetime in Hawking's original calculation, while accounting for back-reaction effects. We argue that the negative-energy Hawking modes need to be  regularly integrated out; and so these are effectively subsumed by the BH and, as a result, the number of coherent negative-energy modes $N_{coh}$ at any given time is parametrically smaller than the total number of the Hawking particles $N_{total}$ emitted during the lifetime of the BH. We find that $N_{coh}$  is determined by the width of the  BH wavefunction and scales as the square root of the BH entropy. We also find that the coherent negative-energy modes are strongly entangled with their positive-energy partners.  Previously, w
 e have found that $N_{coh}$ is also the number of coherent outgoing  particles and that information can  be continually transferred to the outgoing radiation at a rate set by $N_{coh}$.  Our current results show that, while the BH is semiclassical, information can be released  without jeopardizing the nearly maximal inside-out entanglement and imply that the state of matter near the horizon is approximately the vacuum. The BH firewall proposal, on the other hand, is that the state of matter near the horizon deviates substantially from the vacuum, starting at the Page time. We find that, under the usual assumptions for justifying the formation of a firewall,  one does indeed form at the Page time.  However, the possible loophole lies in the implicit assumption that the number of strongly entangled pairs can be of the same order of $N_{total}$.

\end{abstract}
\newpage
\renewcommand{\baselinestretch}{1.5}\normalsize

\section{Introduction}

That a black hole (BH) emits thermal radiation \cite{Hawk} presents the following puzzle: How does an initially pure state of matter evolve into a mixed state of radiation without violating the principles of quantum mechanics? This is, in a nutshell,  the BH information-loss paradox. (See \cite{info} for Hawking's seminal discussion and \cite{info2,info3,info4} for reviews.)

Although this puzzle is regarded by many as an  open question, most of the recent  attention in this context has gone to a related issue that is known as the ``firewall'' paradox \cite{AMPS}. Also see \cite{Sunny1,info4,Braun,Mathur} for earlier versions of the same  idea and \cite{fw1,fw2,fw3,fw4,Sunny2,avery,lowe,vv,pap,AMPSS,lowefw2,SMfw,pagefw,VRfw,MP,bousso2,newfw1,newfw2,newfw3} for what is just a sample of the ensuing discussion.  From this new perspective, one assumes that the radiation does purify eventually and then asks what are the consequences to the standard picture of an observer falling harmlessly through the horizon. From an inspection of the literature, one finds that the answers range from nothing at all to the observer being set a blaze in a sea of high-energy quanta. Obviously, controversy abounds.

A simplified account of the firewall problem goes as follows: Let us parse the BH radiation  into  three  subsystems; the ``early'' Hawking particles,  the ``late'' Hawking particles and the interior ``partners'' of the late Hawking modes. Early and late in this context means before or after the so-called Page time \cite{page}, which is the midway point of entropy transfer. Following many, let us call the subsystems  $A$, $B$ and $C$ respectively. Now, for  the radiation to purify, $A$ and $B$ must be  highly entangled. However, for an observer to fall through the horizon without trauma, $B$ must be close to maximally entangled with $C$. But this is a contradiction because of the monogamy-of-entanglement rule; no system can be simultaneously highly entangled with two different systems. And so, given that the purification of the radiation is true, $B$ and $C$  cannot be maximally entangled. Hence, the state of the near-horizon radiation differs substantially from the vacuum st
 ate and, therefore, the horizon must be a highly excited region that is filled with  non-partnered quanta.  As a consequence, the free-falling observer can expect to burn up on route or, put metaphorically, encounter a firewall.

 See \cite{bousso2} for a recent clarification of the proposal  and for an additional discussion on monogamy of entanglement and purity in this context.

There has been a variety proposals for circumventing  the firewall paradox, many of which have focused on the explicit assumptions in \cite{AMPS}, which are based on the proposed tenets \cite{sussplus} of BH complementarity \cite{hooft} and have already been countered by the original authors \cite{AMPSS,MP}.
A relatively new development is the issue of state dependence in
some of  the proposed
complementarity maps; see \cite{bousso2,newfw1,newfw2} in particular.

We would like to point out an implicit assumption that is being made by both the original paper  and in  many subsequent articles: Namely, that the number of paired Hawking modes is about the same as the  total number of emitted Hawking particles up until (at least) the Page time.  This assumption was  made by Page in his original quantum-information treatment of a radiating  system  \cite{page} and is also made in all models for  which the  Page time is the moment when information  becomes accessible.   However, we will argue that, for an evaporating BH, this assumption is not necessarily correct and its consequences should therefore be reconsidered.

The firewall argument  relies on the standard description of BH evaporation as  developed by Hawking \cite{Hawk,info}. The positive-energy Hawking modes and their negative-energy partners are created continuously  throughout the evaporation process and accumulate near the horizon; just outside and just inside the horizon, respectively. The positive-energy modes escape to infinity, where they are  observed as a thermal flux of radiation. The thermal nature of this  radiation was established by Hawking from a direct calculation of the density matrix for the outgoing modes and  does not require any knowledge about the in-going partners. However, from the pair-production perspective, the thermal nature results from tracing over the negative-energy members of the maximally entangled pairs. The pairs are in a thermofield-double state; however, each of the pairs is produced in a process that is independent of  the production of all the others, so that the pairs themselves are incohe
 rent.

Hawking's setup treats the BH as a strictly classical geometry. In \cite{RB}, it was proposed that the BH information paradox originates  from this assumption. (See \cite{Dvali1,Dvali2,Dvali3} for overlapping ideas.)  On the basis of
\cite{RM,RJ}, it was also proposed in \cite{RB} that the leading semiclassical corrections resulting from  quantum  fluctuations of the background geometry should be taken into account by assigning a wavefunction to the BH.  The parameter that  controls the strength of the semiclassical corrections was identified as the ratio of the Compton wavelength of the BH $\;\lambda_{BH}=\hbar/M_{BH}\;$ to its  radial (Schwarzschild) size $R_S$. In \cite{flucyou},  we have proposed a concrete scheme for evaluating the semiclassical corrections using the wavefunction of \cite{RM,RB}. The parameter that controls the strength of the semiclassical corrections was denoted by $C_{BH}$ and  calibrated  more precisely,
$\;C_{BH}=\frac{\lambda_{BH}}{2\pi}/R_S=1/S_{BH}\;$ ($S_{BH}$ is
the Bekenstein--Hawking entropy). This parameter can be viewed as a  dimensionless $\hbar$ that is controlling quantum corrections.

We have, in two recent articles \cite{slowleak,slowburn}, gone on to apply this idea to the calculation of the Hawking radiation. There, Hawking's calculation was repeated by replacing the classical collapsing shell of matter
({\em i.e.}, the incipient BH) with a semiclassical one that is  endowed with a  Gaussian  wavefunction.  The wavefunction introduces a new scale into the problem via its quantum width. The Bohr correspondence principle has  been invoked to show
that this width should be Planckian \cite{RM,RB}.

We have recalculated the density matrix of the outgoing particles and obtained a picture that is similar to Hawking's in the limit $C_{BH}=0$.  But, for a finite value of $C_{BH}$ (albeit, a very small one), our picture differs significantly from that found by Hawking. The  Hawking density matrix for the BH radiation is strictly diagonal, whereas our matrix contains small off-diagonal elements of order $\sqrt{C_{BH}}$  in the same basis. The effect of these elements on the eigenvalues of the matrix is initially small; however, as the number $N$  of emitted particles grows, so does the changes to the eigenvalues.

Another important distinction  is  the degree of coherence of the radiating particles. In Hawking's case, the emitted particles are incoherent at any time. In our picture, the number of coherent particles at any given moment is finite and  set by a scale that we refer to as the radiation coherence time $t_{coh}$. Typically,  $\;t_{coh}=R_S^2/l_p\;$ (where $l_p$ is the Planck length) and the number of coherent  Hawking particles $N_{coh}$ is equal to the number of particles that are emitted over this time scale, $\;N_{coh}=1/\sqrt{C_{BH}}=\sqrt{S_{BH}}\;$. These estimates are accurate during most of the lifetime of the BH and  become inaccurate only  at the  last stages of evaporation.

The appearance of a coherence time scale in our formalism is quite natural
because of   the following reasoning: The back-reaction of the emitted
particles
on the collapsing shell of matter leads to a time-dependent wavefunction.
Let us then consider  the time required for  this wavefunction
to change significantly. An inspection of
its formal expression (see Eq.~(\ref{wave})) indicates that this happens when the Schwarzschild radius shrinks by an amount  $\;\Delta R_S\sim \sqrt{C_{BH}}R_S\sim l_p\;$. Then, since $\;\Delta R_S =\frac{\partial R_S}{\partial t}\Delta t\sim-\frac{l_p^2}{R_S^2}\Delta t\;$, it follows that  $\;\Delta t\sim\frac{R_S^2}{l_p}=t_{coh}\;$. Hence, the coherence time originates as the interval for which the overlap of the wavefunction at different times becomes small. The fact that
$\;N_{coh}\ll S_{BH}\;$ (equivalently, $\;t_{coh}\ll \tau_{BH}\;$, where $\tau_{BH}$ is the BH lifetime) is a consequence of the width of the wavefunction being much smaller than the Schwarzschild radius. This  hierarchy of scales  can be attributed  to the BH being semiclassical, $C_{BH}\ll 1$.

Both differences are directly related to treating the BH as a semiclassical quantum state rather than a classical geometry.  The strength of the off-diagonal terms is the dimensionless width of the wavefunction $\sqrt{C_{BH}}$ and the coherence scale is related to the overlap of the wavefunction at different times. Thus, our explicit calculations strengthen the ideas expressed in \cite{RB} that treating the BH as a semiclassical state is an essential element in resolving many of the issues surrounding BH physics.

Our framework, as described in \cite{slowburn}, incorporates  the back-reaction of the emitted particles on the collapsing shell in addition to the shell's wavefunction. What we have found is that Hawking was correct in dismissing the effect of the back-reaction when the background is strictly classical. However, for our semi-classical framework, the back-reaction  on the shell does become important.

In this paper, we extend our previous calculations to the negative-energy
modes and  try  to learn about the implications of our semiclassical framework on the pair-production perspective of BH evaporation.

From the pair-production perspective, it is appropriate to integrate out the shell of matter \cite{info}. This amounts to replacing the shell by an eternal BH geometry with specific boundary conditions for matter fields in this geometry. In Hawking's model, this replacement is approximately valid for all times. From the pair-production point of view, it looks as if the negative-energy modes  are concurrently being subsumed into and annihilating with the BH. Consequently, the mass of the BH is decaying with time at a rate dictated by the thermal emission of the BH.  When combined, these observations suggest a picture of the negative-energy modes being continually recycled at some approximately constant rate while the total number of positive-energy Hawking particles is  steadily growing at a rate that is set by the thermal emission of the BH.

In our framework,  the replacement of the collapsing shell with the eternal BH geometry has limited validity. The identity of the negative-energy modes is not well defined and is sensitive to the decrease of the BH mass and radius due to the back-reaction.  To resolve this issue of mode identities, we propose that, at regular intervals whose duration is one
interval of coherence time, the shell spacetime has to be replaced by an eternal BH of smaller mass and smaller horizon radius.

The repetition of the process of integrating out the shell requires us to reassign the wavefunction to the BH and to redefine the pair basis  accordingly.  It follows that the negative-energy modes should be traced over at these regular intervals. The positive-energy  partners of the negative-energy modes  that have been traced over then become part of the state of the external radiation.

It may appear that the process of regularly tracing over the negative-energy modes will lead to information loss.  However, these anti-particles have actually been absorbed into the interior matter, and so the information about these  modes is not lost but rather stored inside the BH.  In Subsection~7.3, we recall a
qualitative discussion whose  aim is to explain how the information could nevertheless be retrieved towards the end of the evaporation. For now, let us  discuss an alternative point of view that will be elaborated on, quantitatively, in an upcoming article \cite{endgame}: As the state of  the BH is the purifier for the outgoing Hawking radiation, someone who is continuously monitoring the external radiation would know about the state of the BH, including the subsumed anti-particles, as it evolves in time. Such an observer would then conclude that the radiation is monotonically  purifying as the evaporation proceeds.

The main objective of the current paper is to substantiate in a quantitative way the above description of BH evaporation in terms of the Hawking pairs.  We calculate the quantum state of the in--out sector and  show that the relevant pairs are in a state of nearly maximal entanglement, at least until the late stages of the evaporation process. But this is all that is  needed because we have  already shown that the early and late Hawking modes  attain  almost full entanglement
but only  at a similarly  late time \cite{slowleak,slowburn}.

The plan for the   rest of the paper goes as follows:
The subsequent section begins with a brief review of our  previous results \cite{slowleak,slowburn}. In Section~3, we reformulate our semiclassical density matrix for external radiation into a quantum state that  describes the Hawking in--out modes.
Next, in Section~4, we elaborate on the above ideas about modeling the back-reaction and the need to
trace out the negative-energy modes at regular intervals.
The consequences of this model for the BH Hilbert space in the pair basis is the topic of Section~5.
We then construct a multi-pair density matrix in Section~6 as a prelude
to determining  the entanglement entropy of the in--out sector.
The latter calculation  is carried out in Section~7.
The paper concludes with a brief summary and discussion in Section~8.

\subsection{The case against firewalls}

Before proceeding, let us give a brief account of why
our framework is able to evade the firewall problem without invoking changes to the standard rules of quantum mechanics.

As already  mentioned, there is a hidden assumption in the current literature on firewalls that almost all  of the produced in-modes are strongly entangled with their out-mode partners up until (at least) the Page time. After this, the external radiation is entropically dominant over the  interior subsystem, which consists of the BH interior including the  negative-energy modes; and so the status of these in-modes becomes a moot point. This is quite clear from the  analysis of Page \cite{page} (also see \cite{HaydenPreskill}). In the total BH--radiation system, the larger of the two  subsystems  holds most of the system's entanglement in the form of  internal correlations, rather than as a mutual entanglement with the smaller one. This is the significance of the Page time; the moment that the interior region and exterior particles exchange their previous roles as the dominant and submissive subsystem.

One can now see why a firewall is inevitable for the orthodox  picture of evaporation. After the Page time, most of the entanglement is necessarily stored in correlations between early and late Hawking particles, these being the  constituents of the dominant exterior subsystem. The now ``unpartnered''  in-modes will make the horizon a dangerous place.

But one can also recognize a possible loophole for evading the firewall problem. First consider  that semiclassical deviations from maximal entanglement of the in-out pairs are controlled by the number of entangled in--out pairs. The deviations of the near-horizon state from the vacuum must therefore be controlled by this same number. Now suppose  that the number of pairs is parametrically smaller than the total number of emitted Hawking  particles. If so, then both the rate of information release and the degree of in--out entanglement will be controlled by the number of pairs rather than the size of the subsystems. What we have found is  that the number of entangled pairs is equal to $N_{coh}$, which is indeed parametrically smaller than the total number of emitted particles. If this possibility is realized, then there must be another component that purifies the outgoing radiation during most the lifetime of the BH, otherwise
unitarity will not be preserved.
In our model, this component is the collapsed matter as represented by the shell and its wavefunction.

Because information is being  released as the outgoing radiation purifies, there will come a time when the rate of information release is too large for the (nearly) maximal  in--out  entanglement to  be maintained. At this point, assuming standard quantum mechanics,  one could expect some large deviation of the near-horizon state from the vacuum and for the associated  firewall to  appear. What we find is that the new tipping point occurs parametrically close to the end of the evaporation, one interval of the coherence time before the BH totally evaporates. In \cite{slowburn}, this is what we have called the transparency time $t_{trans}$. So that the Page time has, in effect, been moved to a time $t_{trans}$ that is late in the evaporation process. But, by this time, the BH can no longer be considered as semiclassical and there is no longer any good reason to  expect its  horizon to be a serene  place (see below). How different is this late-time near-horizon state from the va
 cuum  and what are its properties   are interesting questions that we intend to answer in the future \cite{noburn}.

Let us explain  why, at times $t>t_{trans}$, the BH can  no longer
be considered  semiclassical even though it can still be macroscopically large
with a near-horizon curvature that is small in Planck units. Our basic claim is that, for $t>t_{trans}$, an evaporating BH lacks a  semiclassical description, irrespective of its size or the smallness of the curvature.
The essential point is that  the transparency time coincides with the time when $\;N_{coh}C_{BH}\simeq 1\;$ \cite{slowburn}, meaning that $\;N_{coh}\simeq S_{BH}\;$. Then, the number of negative-energy particles in the near-horizon region
that  are about to fall into the BH is about $S_{BH}$. It follows that their total energy is equal in magnitude to the energy of the remaining BH.  Such a situation does not correspond anymore to the standard semiclassical picture of a large BH  being weakly  perturbed by a small number of negative-energy modes. Rather, back-reaction effects from the in-falling  modes become large and significant, and so the notion of a  nearly classical geometry for the BH  is no longer tenable.

\section{Review of semiclassical black holes and the radiation density matrix}

Here, we review the results from our previous semiclassical model
of the outgoing BH radiation. The framework was initially constructed in \cite{slowleak} and later improved upon in \cite{slowburn} by  accounting for time-dependent and  back-reaction effects.

\subsection{Conventions}

Our conventions are the same as in \cite{slowburn} and
repeated here for completeness.

Our units are such that Planck's constant $\hbar$, Newton's constant $G$
or the combination $l_p=\sqrt{\hbar G}$  are  explicit and  all other fundamental constants are set to unity.

We assume a four-dimensional Schwarzschild BH  (generalizations to higher dimensions are straightforward) of large but finite mass $\;M_{BH}\gg\sqrt{\hbar/G}\;$,
with the  metric
$\; ds^2 = -\left(1-\frac{R_{S}}{r}\right)dt^2+ \left(1-\frac{R_{S}}{r}\right)^{-1}dr^2
+r^2d\Omega^2_2\;$, where  $\;R_{S}=2GM_{BH}\;$ is the horizon radius.
We use the dimensionless advanced-time coordinate $\; v =\frac{1}{R_S}\left(t+r^{\ast}\right)\;$, where
$\; r^{\ast}=\int^r\;dr \sqrt{-g^{tt}g_{rr}} = r+R_{S}\ln(r-R_{S})\;$. Our frequencies or $\omega$'s are also dimensionless and measured in units of $1/R_S$.

The Hawking temperature $T_H$ and Bekenstein--Hawking entropy $S_{BH}$ of the BH are given by $\;T_{H}=\frac{\hbar}{4\pi R_{S}}\;$
 and  $\;S_{BH}=\frac{\pi R_{S}^2}{\hbar G}\;$.

All classically  evolving quantities ({\em i.e.}, all functions
of $\;R_S=R_S(t)\;$) should be regarded as time dependent.

\subsection{Time-dependent semiclassical radiation density matrix}

The  meaning of the semiclassical density matrix, $\;\rho_{SC}=\rho_{H}+\Delta\rho_{SC}\;$, is the following:
\be
\rho_{SC}(\omega,\wt{\omega}~;C_{BH})\;=\; \langle \Psi_{shell}(v_{shell})| \rho(\omega,\wt{\omega}) | \Psi_{shell}(v_{shell})\rangle\;,
\label{SCDM}
\ee
where $\Psi_{shell}(v_{shell})$ is the wavefunction for the collapsing shell of matter. The diagonal Hawking density matrix $\rho_H$ picks up a correction  $\;\Delta\rho_{SC}\sim\sqrt{C_{BH}}\;$ that  introduces an  off-diagonal modification.

We  find $\Psi_{shell}(v_{shell})$
by starting with the wavefunction for the $S$-mode of a  Schwarzschild BH
in Einstein gravity \cite{RM,RB,flucyou} and then assume that this describes the  wavefunction of the shell
in the limit of horizon formation, $\;R_{shell}\to R_{S}\;$ ($R_{shell}$ is the shell's radius).
 This leads to
\be
\left.\Psi_{shell}(R_{shell})\right|_{R_{shell}\to R_{S}}\; = \;{\cal N}^{-1/2} e^{-\frac{\left({R}_{shell}-R_{S}\right)^2}{2C_{BH} R_{S}^2}}
\label{wave}
\;,
\ee
where  ${\cal N}$ is a normalization constant and
$\;C_{BH}=S_{BH}^{-1}\;$ is the aforementioned  classicality parameter.
During most of the lifetime of the BH, when $\;C_{BH}\ll 1\;$, the spacetime
can be treated as classical up to corrections going as a power series in $C_{BH}$.
The correction $\Delta\rho_{SC}$ contains a factor $C^{1/2}_{BH}$ and so is suppressed relative to $\rho_{H}$. The classicality parameter $C_{BH}$  increases slowly and  monotonically throughout the lifetime of the BH and trends to order unity when the size of the BH approaches
the Planck scale.

Our prescription for calculating expectation values is
\be
\langle  \wh{O}(v_{shell})\rangle\;=\;
\frac{4\pi}{\N} \int\limits_{-\infty}^\infty dv_{shell}\;
R_{shell}^2(v_{shell})\ e^{-\frac{(v_{shell}-v_0)^2}{C_{BH}}} {O}(v_{shell})\;,
\label{vfunction}
\ee
where $v_0$ is the classical value of $v$ at horizon crossing, $\wh{O}$ is a generic operator and
we have used that
$\;v_0-v\simeq R-R_S\;$ in the near-horizon limit.

Our later analysis in \cite{slowburn} entailed a time-dependent calculation that accounted for
the different shell-crossing times of the Hawking modes and
for the   effect of the back-reaction on the shell. The number of emitted particles was found to be a good time coordinate, and Eq.~(\ref{SCDM}) gets
corrected to
\bea
&&\rho_{SC}(\omega,\wt{\omega};N_T;N^{\prime},N^{\prime\prime})=
\int\limits_{-\infty}^{v_0}dv \int\limits_0^\infty d\omega'\int\limits_0^\infty  d\omega''  \frac{1}{2\pi} e^{iv(\omega^{\prime}-\omega^{\prime\prime})} \cr &\times& e^{i\omega^\prime\left( v_{shell}(N_T)-v_{shell}(N^\prime)\right)  }
e^{-i\omega^{\prime\prime}\left( v_{shell}(N_T)-v_{shell}(N^{\prime\prime})\right) }
\cr &\times& \langle\Psi_{shell}(v_{shell}(N_T))| \beta^{\ast}_{\omega^{\prime}\omega,~SC}(N_T)
\beta_{\omega^{\prime\prime}\widetilde{\omega},~SC}(N_T)
|\Psi_{shell}(v_{shell}(N_T))\rangle\;. \ \
\label{tdrho}
\eea
where $N_T$ is the ``time'' since the BH formed,
$N$ and $N'$ are the respective shell-crossing times of a given pair of particles and the $\beta$'s are  Bogolubov coefficients.  The subscript $SC$ on the $\beta$'s indicates that
these have been suitably reformulated in terms of the fluctuating parameter $v_{shell}$.

What we have found is that, after all integrations have been performed, the off-diagonal elements of the semiclassical density matrix pick up a time-dependent
``suppression'' factor $D(N_T,N^\prime,N^{\prime\prime})$,
\be
\Delta\rho_{SC}(N_T,N^{\prime},N^{\prime\prime}) \;= \;
D(N_T,N^\prime,N^{\prime\prime}) \Delta\rho_{SC}(C_{BH}(N_T))\;.
\ee
where
\be
D(N_T;N',N'')\;\equiv\;\frac{1}{2}\Big[ e^{-\frac{1}{4} \frac{\left[C_{BH}(N^\prime) (N_T-N^\prime)\right]^2}{ C_{BH}(N_T)}}+e^{-\frac{1}{4}\frac{\left[C_{BH}(N^{\prime\prime}) (N_T-N^{\prime\prime})\right]^2}{ C_{BH}(N_T)}}\Big]\;.
\label{suppfactor}
\ee
An important consequence of this  factor is that our perturbative treatment --- which
formerly broke down at best by the Page time, $\;N_T=\frac{1}{2} S_{BH}(0)\;$  ---
can now be continued until much later in the process;
essentially, until there are only $S^{1/3}_{BH}(0)$ particles
remaining to be emitted (this being the transparency time).

The same suppression factor is obtained in the upcoming  in--out
treatment.
It is still  irrelevant to the Hawking part of the matrix,
which already  carries the implicit suppression $\delta(N'-N^{''})$ because, in
this case, the in- and out-modes emerge as perfectly entangled pairs.

The suppression is insignificant when $\;\Delta N
\equiv N_T-N \leq N_{coh}(N_T;N)\;$, for  which
\be
 N_{coh}(N_T;N) \;\equiv\;  \frac{\sqrt{C_{BH}(N_T)}}{C_{BH}(N)}\;.
\label{coh}
\ee
We call this the coherence time because, for $\;\Delta N > N_{coh}(N_T;N)\;$,
the density matrix elements and, therefore, the particle correlations become highly suppressed.
For most of the lifetime of the BH, $\;C_{BH}(N)\simeq C_{BH}(N_T)\;$, so that
$\;N_{coh}(N_T;N)\simeq C_{BH}^{-1/2}(N_T) \simeq \sqrt{S_{BH}(N_T)}\;$.
At late times, however, this distinction can become important,
as $C_{BH}(N_T)$ is monotonically growing as $\;\partial_{N_T} C_{BH}=C_{BH}^2\;$
and reaches unity for a Planck-sized BH.

\section{Semiclassical state of the Hawking pairs}

Before proceeding with the calculation of the pair semiclassical state, let us describe Hawking's pair-production picture and remark on some caveats.

Hawking's original choice of basis is that of the collapsing-shell model \cite{Hawk}. For this choice, the negative-energy modes defy  an obvious particle interpretation because, as far as an external observer is concerned, the separation between positive and negative energies becomes ambiguous inside of  the shell's horizon. This ambiguity motivated Hawking to choose a different basis in his subsequent  information-loss article \cite{info}. This latter setup  assumes an analytically continued Schwarzschild spacetime ({\em i.e.}, an eternal BH geometry), for which such a separation can be made without ambiguity. In effect, to discuss the negative-energy modes, the shell is integrated out and replaced by an eternal BH geometry with a particular choice of boundary conditions for the matter fields. So that,
in this model, the vicinity of the horizon is devoid of any matter.

In principle, to calculate the semiclassical state for the pair-production model, we would need to know the wavefunction of the BH from an in-falling observer's perspective and then  proceed along the lines of Subsection~2.2. However, as explained below, we bypass this difficulty by exploiting a relationship between the in- and out-modes that allows us
to use the external observer's wavefunction.

Hawking chose to work with the $w_{\omega}$ and $y_{\omega}$ basis of Section~4 in \cite{info},  rather than  the
$q_{\omega}$ and $p_{\omega}$ basis of \cite{Hawk}.  The $w$'s are defined to have zero Cauchy data on $\cal{I}^-$ and on the portion of the past horizon outside the future horizon. They represent particles that are always inside the future horizon. The $y$'s are defined to have zero Cauchy data on $\cal{I}^-$ and on the portion of the past horizon inside the future horizon, as well as having positive energy with respect to the retarded time $u$ on the portion of the past horizon outside the future horizon.  Hawking also showed that, as far as their action on the initial vacuum is concerned, the $y_{\omega}$'s are equivalent to $p_{\omega}$'s and the $w_{\omega}$'s are equivalent to  $p_{\omega}^{\dagger}$'s.

The initial vacuum is defined at $\cal{I}^-$ and the past horizon for the $w$--$y$ basis but only
at ${\cal I}^-$ for the $p$--$q$ basis. However, this distinction is irrelevant  to the geometry of interest, the interior and exterior regions of the future horizon.

Hawking's choice of the $w$--$y$ basis corresponds to a partial tracing over some of the negative-energy modes (see below). We rather need to begin with modes that correspond to the complete ``untraced'' negative-energy modes. The complete horizon
modes are denoted by Hawking as $f_{\omega}^{(3)}$ and $f_{\omega}^{(4)}$.
They are defined as having  zero Cauchy data on $\cal{I}^-$ and, on the whole of the past horizon, they have time dependence of the form   $e^{\pm i \omega u}$, respectively. The operator forms of $f_{\omega}^{(3)}$ and $f_{\omega}^{(4)}$ are given  in Eq.~(4.16) of \cite{info} (re-expressed here
in our notation),
\be
\wh{f}_{\omega}^{(3)} \;=\; \frac{1}{\sqrt{1-c_{\omega}^2}}\left[\wh{y}_{\omega}-c_{\omega}\;
\wh{w}^{\dagger}_{-\omega}\right]\;,
\label{f3}
\ee
\be
\wh{f}^{(4)}_{\omega} \;=\; \frac{1}{\sqrt{1-c_{\omega}^2}}\left[\wh{w}_{-\omega}-c_{\omega}\;
\wh{y}^{\dagger}_{\omega}\right]\;,
\label{f4}
\ee
where $\;c_{\omega}\equiv e^{-2\pi \omega}\;$.

This discussion highlights the fact that the pair-production picture has limited validity and is particularly sensitive to back-reaction induced deviations away from the eternal BH geometry. This sensitivity will be essential in the following.

Our eventual task is to calculate the entanglement between the out-modes --- the incipient Hawking particles ---  and the in-modes --- their negative-energy partners. For this calculation, we will  first require  the matrix elements  for the final  vacuum $|0_+\rangle$  and then  the expectation value of this matrix with respect to the initial  vacuum $|0_-\rangle$. To this end, we will  calculate the in--out analog of the out--out density matrix of our previous studies
\cite{slowleak,slowburn}. However, the resulting matrix $\rho_{in-out}$ should not be viewed as a density matrix but as the coefficients of the terms of an entangled state,
\be
\Psi_{pair}(\omega_{out},-\wt{\omega}_{in})\;=\;\frac{1}{Z} \int d\omega\; d\wt{\omega}\; \rho_{in-out}(\omega_{out},-\wt{\omega}_{in}) |\omega_{out}\rangle |\wt{\omega}_{in}\rangle\;,
\ee
where $Z$ is a normalization factor.

The one-pair matrix that we have in mind is
then the in--out analog of the following
 (with the expectation value implied on the left-hand side):
\be
\rho_{out-out}(\omega,\wt{\omega})  \;=\;
\langle 0_-| \left(\wh{F}^{\dagger}_{\omega} + \wh{F}_{\omega}\right)| 0_+ \rangle
\langle 0_+| \left(\wh{F}^{\dagger}_{\wt{\omega}}
+ \wh{F}_{\wt{\omega}}
\right)| 0_- \rangle  \;.
\label{denmat}
\ee
Here,  our notation  is such that   $\;\wh{F}_{\omega}= F_{\omega}\left(u\right)\wh{a}_{\omega}(F)\;$ includes both the wavefunction of the out-mode as a function of retarded time $\;u=u(v)\;$
and the  annihilation operator. Analogous forms
for other hatted modes  are used below.
We are  currently considering a fixed value of advanced time $v$,
 but this coordinate is later integrated out.

It is a difficult task to calculate $\rho_{in-out}$ directly. As already stated,  we would need to know the wavefunction of the horizon from an in-falling observer's perspective in the eternal BH geometry. Rather than doing this, we will express the in-modes in terms of a linear combination of the out-modes. We will  then evaluate the corresponding $\rho_{out-out}$ and use the result to find
$\rho_{in-out}$.

For the out--out case, one finds that the only relevant contribution of the four terms is  $\wh{F}_{\omega}^{\dagger}\wh{F}_{\widetilde{\omega}}$, which leads
to
\bea
\rho_{out-out}(\omega,\wt{\omega})  \;= \;
\sum\limits_{\omega',\omega''}
f_{\omega'}(v) \beta^{\ast}_{\omega',\omega} \beta_{\omega'',\wt{\omega}}
\ f_{\omega''}^{\ast}(v)
\;,
\eea
where the $\beta$'s are the ``negative-energy'' Bogolubov coefficients and $f_{\omega'}(v)$
is a basis function for the initial vacuum.
The ``positive-energy'' Bogolubov coefficients or  $\alpha$'s enter through the
other terms  and could contribute
in principle. However,  as explained in Subsection~2.4 of \cite{slowleak}, these end up to be irrelevant for particle production for our semiclassical
analysis just like in  Hawking's treatment \cite{Hawk}.

Our out-modes are related to Hawking's modes in \cite{info} as follows:
\be
\wh{F}_\omega\;=\; t_\omega \wh{y}_\omega+r_\omega \wh{z}_{\omega}\;,
\ee
where $t_\omega$ and $r_\omega$ are the transmission and reflection coefficients, respectively.
The mode $\wh{z}_\omega$ is irrelevant for the pair-production process.
Hence, for the purpose of calculating the in--out matrix, we can equate
\be
\wh{y}_\omega \;=\; \frac{1}{t_\omega} \wh{F}_\omega\;.
\ee

To obtain the correct  set of in- and out-modes for current purposes, we
recall the following  identity from \cite{info}:
\be
\wh{w}^{\dagger}_{\ww}|0_-\rangle \;=\; c_{\omega}^{-1}\; \wh{y}_{\omega}|0_-\rangle\;,
\label{in-op}
\ee
where all   frequencies are assigned
according to the perspective of an external observer
({\em i.e.}, $\;\omega>0\;$ in all cases).
In other words, the creation of a negative-energy excitation is
equivalent to the annihilation of a positive-energy one with the same
magnitude of energy.

We propose that the correct definition for a complete horizon mode
is as follows:
\be
\wh{W}_{\omega}\;=\;\frac{1}{2}
\left[\wh{f}^{(3)}_{\omega} + (\wh{f}^{(3)}_{\omega})^{\dagger} \;+\;
\wh{f}^{(4)}_{\omega} + (\wh{f}^{(4)}_{\omega})^{\dagger}\right]\;,
\ee
where it is understood that only its negative-energy component
contributes to the in--out density matrix as depicted in
Eq.~(\ref{inoutDM}) below.
Then, using  Eqs.~(\ref{f3},\ref{f4}) as well as  Eq.~(\ref{in-op}) to trade off $\wh{w}_{\omega}$'s for $\wh{y}_{\omega}$'s, we can express $\wh{W}_{\omega}$ in terms of the out-mode $\wh{y}_{\omega}$,
\bea
\wh{W}_{\omega}&=&  \frac{1}{2} \frac{1}{\sqrt{1-c_{\omega}^2}}\left(1-c_{\omega}\right)\left(1+
\frac{1}{c_{\omega}}\right)\left[\wh{y}_{\omega}+\wh{y}_{\omega}^{\dagger}\right]\nonumber\\
&=& \frac{1}{2}\frac{\sqrt{1-c_{\omega}^2}}{c_{\omega}}\left[\wh{y}_{\omega}+ \wh{y}^{\dagger}_{\omega}\right]
\;.
\eea

The in--out matrix has now  been expressed entirely in terms of out-modes,
\bea
\rho_{\rm in-out}(\omega,-\wt\omega)&=& \frac{1}{2}
\langle 0_-|(\wh y_\omega^{\dagger}+ \wh y_\omega)|0_+\rangle
\langle 0_+|(\wh W_{\wt{\omega}}^{\dagger}+\wh W_{\wt{\omega}})|0_-\rangle
\nonumber \\
&=& \frac{\sqrt{1-c^2_{\wt\omega}}}{c_{\wt\omega}}\langle 0_-|(\wh y_\omega^{\dagger}+ \wh y_\omega)|0_+\rangle
\langle 0_+|(\wh y_{\wt\omega} +\wh y^{\dagger}_{\wt\omega})|0_-\rangle
\label{inoutDM} \\
&=&  \frac{\sqrt{1-c^2_{\wt\omega}}}{c_{\wt\omega}}\langle 0_-|(\frac{1}{t_{\omega}^{\ast}}\wh F_\omega^{\dagger}+
\frac{1}{t_{\omega}} \wh F_\omega)|0_+\rangle
\langle 0_+|(\frac{1}{t_{\wt\omega}}\wh F_{\wt\omega}
+\frac{1}{t_{\wt\omega}^{\ast}}\wh F^{\dagger}_{\wt\omega})|0_-\rangle
\;. \nonumber
\eea
Here, the operator $\wh F^{\dagger}_\omega$ should be regarded as
an  excitation of
a horizon mode and not that of an  asymptotic Hawking particle.

Now, just as for the out--out case,  the only contribution to
the density matrix comes from the pair
$\wh F_\omega^{\dagger}\wh F_{\wt\omega}$, as the rest have either rapidly oscillating phases or represent irrelevant non-propagating modes.
It follows that
\be
\rho_{\rm in-out}(\omega,-\wt{\omega})\;=\; \frac{\sqrt{1-c^2_{\wt\omega}}}{c_{\wt\omega}}
\langle 0_-|\frac{1}{t^{\ast}_\omega}\wh F_\omega^{\dagger}|0_+\rangle
\langle 0_+|\frac{1}{t_{\wt\omega}}\wh F_{\wt\omega}|0_-\rangle \;.
\label{rhoinout1}
\ee

Next, expanding the matrix  in terms of
 the basis kets $| f_{\omega'} \rangle$, we obtain
\bea
\rho_{\rm in-out}(\omega,-\wt{\omega} )  &=&
\frac{1}{t^{\ast}_{\omega}t_{\wt\omega}}\frac{\sqrt{1-c^2_{\wt\omega}}}
{c_{\wt\omega}}
 \sum\limits_{\omega',\omega''}
\langle 0_-|f_{\omega'}\rangle\langle f_{\omega'}| \wh{F}^{\dagger}_{\omega}| 0_+\rangle
\langle 0_+| \wh{F}_{\wt{\omega}}|f_{\omega''}\rangle\langle f_{\omega''}|0_-\rangle
\nonumber \\
&=& \frac{1}{t^{\ast}_{\omega}t_{\wt\omega}}\frac{\sqrt{1-c^2_{\wt\omega}}}
{c_{\wt\omega}}\sum\limits_{\omega',\omega''}
f_{\omega'}(v)\langle f_{\omega'}| F_{\omega} \rangle
\langle  F_{\wt{\omega}}|f_{\omega''}\rangle f^{\ast}_{\omega''}(v)
\;,
\label{eqinout}
\eea
where  $|F_{\omega}\rangle$ means
a one-particle ket.

It is the amplitudes in the last line that describe the overlap between
particle modes and basis vectors and, therefore, represent the  Bogolubov coefficients. Because the right-hand side  of
Eq.~(\ref{eqinout}) involves only out-modes, these coefficients are the same as those obtained in the out--out case.
Hence,
$\;\langle f_{\omega'}| F_{\omega} \rangle = \beta^{\ast}_{\omega',\omega}\;$,
$\;\langle  F_{\wt{\omega}}|f_{\omega''}\rangle
= \beta_{\omega'',\wt{\omega}}\;$ and, consequently,
\bea
\rho_{in-out}(\omega,-\wt{\omega})  \;=\; \frac{1}{t^{\ast}_{\omega}t_{\wt\omega}}
\frac{\sqrt{1-c^2_{\wt\omega}}}{c_{\wt\omega}}\;\sum\limits_{\omega',\omega''}
f_{\omega'}(v) \beta^{\ast}_{\omega',\omega} \beta_{\omega'',\wt{\omega}}
 f^{\ast}_{\omega''}(v)
\;.
\eea

We have then ended up with a matrix that is  similar in form to the  matrix for the out--out case.  Again, there is the  Hawking classical-background contribution, except
that it is now   describing maximally entangled pure state of pairs and will lead to a  thermal reduced density matrix, as explained below. It differs from the usual form by a factor which will turn out to be very significant,
\bea
\left[\rho_{\rm in-out}(\omega, -\widetilde{\omega})\right]_H
 &=&
\frac{\sqrt{1-c^2_{\omega}}}{c_{\omega}}\frac{1}{e^{\frac{\hbar\omega}{T_{H}}}-1} \delta(\omega-\widetilde{\omega})\;,
\label{spec}
 \eea
where  $\;T_H/\hbar=(4\pi)^{-1}\;$ is the  dimensionless Hawking temperature.

One formal difference between Eq.~(\ref{spec}) and
Hawking's out--out matrix is that the transmission amplitudes of the out-modes
 through the gravitational barrier no longer appear. This is
sensible because these are horizon modes and not the asymptotically
transmitted  Hawking particles.

 A more important distinction is, however,  the
extra factor of
\be
\frac{\sqrt{1-c^2_{\omega}}}{c_{\omega}}
\;= \;\frac{\sqrt{1-e^{-4\pi\omega}}}{e^{-2\pi\omega}}
\;=\;\sqrt{e^{4\pi\omega}-1}\;=\;\sqrt{e^{\frac{\hbar\omega}{T_H}}-1}\;.
\ee
In this way, we actually end up with the ``square root'' of the Hawking thermal
form,
\be
\left[\rho_{\rm in-out}(\omega, -\widetilde{\omega})\right]_H
\;=\;
\rho^{1/2}_H(\omega,-\wt{\omega})
 \;\equiv\;
\frac{1}{\sqrt{e^{\frac{\hbar\omega}{T_{H}}}-1}} \delta(\omega-\widetilde{\omega})\;.
\label{specSQ}
 \ee

Despite appearances, the  matrix in Eq.~(\ref{specSQ}) is not a  density matrix but, rather,  represents a pure state. The state is a superposition of pairs with weights $\frac{1}{\sqrt{e^{\frac{\hbar\omega}{T_{H}}}-1}}$ and, therefore, a thermofield-double state for the pairs of positive- and negative-energy modes,
\be
|\Psi_{pair,\;H}(\omega_{out},-\wt{\omega}_{in})\rangle\; =\; \frac{1}{Z} \int d\omega\; \frac{1}{\sqrt{e^{\frac{\hbar \omega}{T_{H}}}-1}} |\omega_{out}\rangle |\omega_{in}\rangle\;.
\label{tfdrho}
\ee
Here, $|\omega_{out}\rangle$ denotes a positive-energy out-mode with frequency $\omega$ and $|\omega_{in}\rangle$,  a negative-energy in-mode with frequency $\omega$.  The normalization factor $Z$ will be specified later on.

The full Hawking density matrix is given by
\be
\rho_{pair\;,H}(\omega_{out},-\wt{\omega}_{in},\omega'_{out},-\wt{\omega}'_{in}) \;=\; |\Psi_{pair\;,H}(\omega_{out},
-\wt{\omega}_{in}) \rangle \langle\Psi_{pair\;,H}(\omega'_{out},-\wt{\omega}'_{in}) |
\;.
\ee
 The reduced  matrix for the out-modes is obtained by tracing over the in-modes. So that, as standard for a thermofield-double state, the reduced matrix  goes as the square of Eq.~(\ref{specSQ}) and the correct thermal matrix is indeed obtained.

Meanwhile,
the same lengthy
calculation  as in \cite{slowleak,slowburn} will lead us to the semiclassical correction to the Hawking state,
\be
\Delta\rho_{SC}(\omega,-\wt{\omega};N_T;N^{\prime},N^{\prime\prime})
\;=\  D(N_T;N^{\prime},N^{\prime\prime})
\;C^{1/2}_{BH}(N_T)\Delta\rho_{OD}(\omega,-\wt{\omega}~;N_T)\;,
\label{rhoscf0}
\ee
with
\bea
&&\Delta\rho_{OD}(\omega,-\wt{\omega}~;N_T) \;=\;
\frac{1}{(2\pi)^3} \frac{2}{(\omega \wt{\omega})^{1/2}}
\left(\frac{C_{BH}(N_T)}{4}\right)^{+2i (\omega-\wt{\omega})} \cr &\times&
\frac{\sqrt{e^{4\pi\omega}-1}+\sqrt{e^{4\pi\wt{\omega}}-1}}{2} \Gamma\left(1+2i\omega\right)
\Gamma(1-2i\wt{\omega})
\ \hbox{\Large\em e}^{-\pi(\omega+\wt{\omega})} \
\Gamma\left(\frac{1}{2}- i(\omega-\wt{\omega})\right)\cr
&\times&\Biggl\{
\Gamma \left(2i (\omega-\wt{\omega})\right)\left[
\frac{\Gamma \left(\frac{1}{2}+2i\wt{\omega}\right)}
{\Gamma \left(\frac{1}{2}+2i \omega\right)}
+
\frac{\Gamma \left(\frac{1}{2}-2i\omega\right)}
{\Gamma \left(\frac{1}{2}-2i \wt{\omega}\right)}
\right] + \frac{ i }{\omega-\wt{\omega}}\Biggr\}\;,
\label{rhoscf1}
\eea
where we have symmetrized over the frequencies
 and have subtracted off a diagonal piece
with the understanding that this acts as a small correction to the Hawking
part of the  matrix.
The total  number of  so-far produced pairs $N_T$ is now keeping track of the evolution time and $N'$, $N''$ are the pair-production times.

Equation (\ref{rhoscf0}) should be interpreted as a correction to the thermofield-double state of Eq.~(\ref{tfdrho}). The correction means that  the positive- and negative-energy modes are not exactly maximally entangled, with
$C_{BH}^{1/2}$ controlling the deviation from maximal entanglement. This will be discussed in detail in Sect.~6.

\section{Model of semiclassical back-reaction}

In the previous section, we have highlighted the fact that the particle-pair picture requires one to integrate out the collapsing shell and use the geometry of the eternal BH with appropriate boundary conditions for matter fields. We have also emphasized that, as a consequence,
the pair-production picture has limited validity and is particularly sensitive to deviations away from the  eternal BH spacetime.
We would now like to discuss this issue of  validity  in a more quantitative way and,  in particular,  determine the duration for which the
eternal BH geometry is a good approximation to
the  collapsing-shell model. We will argue that this duration is $t_{coh}$.

The issue of validity of the pair-production picture was not discussed in a meaningful way by Hawking because, in his calculation, the coherence time scale did not appear. Each pair emission was considered to be completely independent of  the previous pairs. Hence, Hawking's choice of the eternal BH geometry had an exponentially small effect.

In the Introduction, we have briefly  outlined a simple model for the back-reaction of the emitted particles when considering the pair-production picture. The basic idea is that  the negative-energy members of the pairs should be regularly traced out as the mass of the BH decreases with time at a rate dictated by the thermal emission. This simple model can be made more precise as follows.

Let us consider the perspective of an external, stationary observer. Then, during one coherence time $t_{coh}$, the BH emits $\;N_{coh}\sim S^{1/2}_{BH}\;$  Hawking particles and its energy decreases by $\;\Delta E_{shell}\sim N_{coh} T_H \sim M_p\;$, where $M_p=l_p^{-1}$ is the Planck mass. The radius of the BH will then shrink by $l_p$, and  its wavefunction becomes much different than it was at  earlier times. This restricts considerations to time intervals of duration $\;\Delta t < t_{coh}\;$.

Over this time interval, the BH and the pair-produced  particles, both positive- and negative-energy ones, are  coherent. But, for time intervals in excess of $t_{coh}$, the negative-energy particles should be traced over (see below), leaving an almost thermal (reduced) density matrix for a block of $N_{coh}$ positive-energy particles with  some small corrections of order $\sqrt{\hbar}$. After their negative-energy partners have disappeared, these should be regarded as   emitted Hawking particles and, as such, will become part of the  out--out radiation density matrix.

Our model for the  back-reaction is quite simple and still needs to be  improved  by providing a more precise description of the interaction of the negative-energy modes with the BH. However, even at the current level of precision, it is already clear that the negative-energy particles cannot keep their identity after an elapse of time  $t_{coh}$. This can be seen from the following argument.

Let us first recall of  the  form of the   out-mode wavefunctions
(with $\omega$, $u$, $v$ dimensionless),
\be
F_{\omega}(u(v))\;\sim\; e^{i\omega u}\;\sim\; e^{i\omega \ln{(v_0-v)}}\;,
\label{outmode}
\ee
and similarly for the in-modes but with the argument in the logarithm
reversed.
Now consider that the positive-energy particles accumulate near the horizon but only on the outside, whereas the negative-energy particles accumulate
on the inside. In Hawking's description, it is not stated how close these modes are to the horizon, so that the distance
$\;\Delta v=v_0-v\;$
remains unspecified. But our case is different because of the uncertainty due to the
quantum  width of the wavefunction. The wavefunction has Planckian width, which implies that  the width of  either particle  layer is about
 $\;\Delta v =l_p/R_S\;$
(or $l_p$ in  dimensional units). Our model  also keeps track of
the shrinking of the  Schwarzschild radius  or,  equivalently,
the decreasing value of $v_0$. After one coherence time,  $v_0$  decreases by an
amount of the same order, $\;\Delta v_0\sim l_p/R_S\;$.

For a time interval $\;\Delta t>t_{coh}\;$,
$v_0$ will have  changed by an amount that is greater in magnitude than
the width of the particle layers,
$\;\Delta v_0 > \Delta v\;$. At this point, the  argument of the
logarithm in
Eq.~(\ref{outmode}) and its in-mode analogue
are likely to change sign.
When such a sign flip does occur, it essentially  exchanges the meaning of the
mode from a positive to a negative-energy excitation  (or {\em vice versa}).
This is
just like
 what would happen if a Rindler  mode is switched from the right wedge
 to the left wedge  of Rindler space. Time flows in the opposite direction in
the left wedge, and so  positive energies become negative. Meaning that, after
the elapse of a coherence time, the splitting into positive and negative energies becomes ill defined and the identity of the near-horizon modes becomes uncertain.

To resolve this issue of mode identities, we propose that, after each coherence time, the eternal BH spacetime should  be reset to a new eternal BH spacetime corresponding to the updated Schwarzschild radius
$\;R_S(t+t_{coh})\simeq R_S(t)-l_p\;$. The pair basis  has to be
 redefined accordingly   and the whole process repeats itself after
the elapse of the  next interval of coherent time.

In general, this limitation on the use of  the eternal BH geometry should be imposed
when the back-reaction is taken into account,
irrespective of  whether  the geometry is treated as classical or
semiclassical.
The only situation in which the negative-energy particles can preserve their identity is for a  truly eternal BH geometry. However, if the coherence time vanishes (as it does for Hawking's model), it is not important that these modes preserve their identity for the purpose of calculating the in--out density matrix.

\section{The black hole Hilbert space}

Let us now discuss how  our previous  model
for the  back-reaction is relevant  to  the structure of the BH Hilbert space ${\cal H}_{BH}$ in the pair basis.

The Hilbert space of an evaporating BH  will approximately factorize into two  Hilbert spaces,  $\;{\cal H}_{BH}\sim {\cal H}_{int}\otimes{\cal H}_{rad}\;$.
Here, ${\cal H}_{int}$ describes
the state of the collapsed matter plus the anti-particles  and
${\cal H}_{rad}$ describes
the outgoing Hawking radiation. The radiation in--out entangled sector is then the boundary or overlap between these two sub-Hilbert spaces,
$\;{\cal H}_{in-out}={\cal H}_{int}\bigcap {\cal H}_{rad}\;$. We know from the analysis of Page that this overlap is small compared to $S_{BH}$ because most of the entanglement is stored as internal correlations within the subsystems and not
as correlations between the subsystems.  But how small?

Our formalism suggests a definitive answer to this last question.  As we now know, it is necessary to
trace over the negative-energy modes after a  time scale of $t_{coh}$ or $\;N_{coh}\simeq S^{1/2}_{BH}\;$ in units of
either number of emitted particles or number of produced pairs.
Our explanation is that this effect  is a consequence of  the wavefunction decohering over the same extent of time. The coherence scale  is then  the
span of time over which it still makes sense to talk about entangled partners; meaning that $\;N_{coh}\simeq S^{1/2}_{BH}\;$ is the typical lifetime of a partnership. This leads to a revision of the orthodox picture: Partnerships are being regularly dissolved  and recycled as the Hawking process of mode creation goes on \cite{Hawk}, with the  total number of entangled partners scaling  as $N_{coh}$. In short, we are arguing that $\;{\rm dim}\left[{\cal H}_{in-out}\right]=N_{coh}\;$.

This recycling process, over the time scale  $\;N_{coh}\ll S_{BH}\;$, provides the means for maintaining the in--out entanglement while information is flowing out of the BH. That the entanglement is   maintained will be clarified in the upcoming analysis, but then what is  the mechanism for information transfer?  It is the wavefunction $\Psi_{BH}$ that plays the  role of  conduit. As the in-modes are  traced out and effectively subsumed into the BH interior, the Schwarzschild radius $\;R_S=R_S(N_T)\;$ decreases and, in turn, induces an evolving value for the coherence scale $\;N_{coh}=N_{coh}(N_T;N)\;$.

An external stationary observer  has direct access to ${\cal H}_{rad}$, whereas a  free-falling observer is able to probe ${\cal H}_{int}$ but at the cost of relinquishing knowledge about the exterior system ${\cal H}_{rad}$. If such observers wish to compare measurements, their only common ground is that of the   boundary region ${\cal H}_{in-out}$. We would like to suggest that this could be a starting point for a definition of BH complementarity \cite{sussplus,hooft} that can survive the firewall paradox.

\section{Multi-Pair density matrix}

To monitor the entanglement of the produced pairs ---  which is the subject of
Section~7 --- it is first necessary to construct a multi-pair
density  matrix for the in--out sector. In light of the previous two sections, it is clear that the multi-pair
matrix should involve $N_{coh}$ pairs of particles.
 Then, following  our  earlier investigations \cite{slowleak,slowburn}, the multi-pair matrix is a  $2 N_{coh}\times 2 N_{coh}$ matrix such that  each  entry
is  a  block with the same dimensionality  in frequency space as the one-pair matrix,
$\;\rho_{SC}(\omega,-\wt{\omega}~;N_T;N',N'')=\rho^{1/2}_{H}(\omega,-\wt{\omega})
+\Delta\rho_{SC}(\omega,-\wt{\omega}~;N_T;N',N'')\;$.
Recall that $\rho^{1/2}_{H}(\omega,-\wt{\omega})$ is defined in Eq.~(\ref{specSQ}) and
the correction  in Eqs.~(\ref{rhoscf0}--\ref{rhoscf1}).
The suppression factor $D(N_T;N^\prime,N^{\prime\prime})$ that appears in $\Delta\rho_{SC}$ is defined in Eq.~(\ref{suppfactor}) and can be re-expressed in a convenient way,
\be
D(N_T;N',N'')\;=\;\frac{1}{2}\Big[ e^{-\frac{1}{4} \frac{(N_T-N^\prime)^2}{ N^2_{coh}(N_T;N^\prime)}}+e^{-\frac{1}{4}\frac{(N_T-N^{\prime\prime})^2}{ N^2_{coh}(N_T;N^{\prime\prime})}}\Big]\;.
\label{suppfactor2}
\ee
As discussed in \cite{slowleak,slowburn}, one can  expect each  entry to pick up a  phase factor $e^{i\theta_{N^\prime,N^{\prime\prime}}}$ ($\;\theta_{N^\prime,N^{\prime\prime}}=-\theta_{N^\prime,N^{\prime\prime}}\;$). But these phases are not relevant to our treatment and will be ignored.

We can  express the multi-particle (MP) state $\Psi_{SC}^{MP}(N_T;N',N'')$ in  Dirac  notation (with frequency
labels now suppressed),
\bea
|\Psi_{SC}^{MP}(N_T;N',N'')\rangle &=& \frac{1}{n_{\ast}}
\rho^{1/2}_H \delta_{N',N''}|N'\rangle | N''\rangle \\
&+&
\frac{C^{1/2}_{BH}(N_T)}{n_{\ast}}\Delta\rho_{{OD}}\;D(N_T;N',N'')
\left[1-\delta_{N',N''}\right]|N'\rangle| N''\rangle\;,\nonumber
\label{nmatrix3}
\eea
where $\;N_T-N_{coh}\lesssim N^{\prime},N^{\prime\prime} \leq N_T\;$. The normalization $n_{\ast}$  will be determined
later on by the requirement that the reduced out-out density matrix be correctly normalized.

The density matrix $\rho_{SC}^{MP}$ corresponding to $|\Psi_{SC}^{MP}\rangle$ is given by the standard expression,
\be
\rho_{SC}^{MP}(N_T;N',N'',N''',N'''')\;=\; |\Psi_{SC}^{MP}(N_T;N',N'')\rangle\langle\Psi_{SC}^{MP}(N_T;N''',N'''')|\;.
\ee

To obtain a reduced density matrix for the out-modes,
we need to re-express the density matrix on the product space $|N_o\rangle\otimes|N_i\rangle$. Then,  $\;\rho_{SC,o\otimes i} =\rho_{SC,o\otimes i}(N_T;N_o',N_i',N_o'',N_i'')\;$,
where the subscripts $i$ and $o$ respectively label in- and out-modes.
This matrix takes the form
\bea
\label{earlylatem1}
&& \hspace*{-0.50in}
 \rho_{SC,o\otimes i}(N_T;N_o',N_i',N_o'',N_i'')
= \frac{1}{n^2_*} \rho^{1/2}_H\otimes\rho^{1/2}_H |N_o'\rangle |N_i'\rangle\langle N_o''|
\langle N_i''|
\left[\delta_{N_o',N_i'} \delta_{N_o'',N_i''}
+ \delta_{N_o',N_i''} \delta_{N_o',N_i''}\right]
\nonumber \\
&+& \frac{C_{BH}(N_T)}{n^2_*}
\;\Delta\rho_{{OD}}\otimes\left(\Delta\rho_{{OD}}\right)^\dagger
\;\times\;\Big\{
\nonumber \\
&& D(N_T;N_o',N_i')D(N_T;N_o'',N_i'')
|N_o'\rangle |N_i'\rangle\langle N_o''|
\langle N_i''|
_{(N_o'\neq N_i'\;,\;N_o''\neq N_i'')}\nonumber \\
&+&
 D(N_T;N_o',N_i'')
 D(N_T;N_o'',N_i')
|N_o'\rangle |N_i'\rangle\langle N_o''|
\langle N_i''|
_{(N_o'\neq N_i''\;,\;N_o''\neq N_i')}
\;\; \Big\}
\nonumber \\
&+&\frac{C^{1/2}_{BH}(N_T)}{n^2_*}
\frac{1}{2}\left[\Delta\rho_{{OD}}\otimes \rho^{1/2}_H
+\rho^{1/2}_H \otimes
\left(\Delta\rho_{{OD}}\right)^\dagger\right]
\;\times\;\Big\{
\nonumber \\
&& D(N_T;N_o',N_i') \delta_{N_o'',N_i''}|N_o'\rangle |N_i'\rangle\langle N_o''|
\langle N_i''|_{N_o'\neq N_i'} \nonumber \\
&+& D(N_T;N_o',N_i'') \delta_{N_o'',N_i'}|N_o'\rangle |N_i'\rangle\langle N_o''|
\langle N_i''|_{N_o'\neq N_i''} \nonumber \\
&+&
 D(N_T;N_o'',N_i') \delta_{N_o',N_i''}|N_o'\rangle |N_i'\rangle\langle N_o''|
\langle N_i''|_{N_o''\neq N_i'} \nonumber \\
&+& D(N_T;N_o'',N_i'') \delta_{N_o',N_i'}
|N_o'\rangle |N_i'\rangle\langle N_o''|\langle N_i''|_{N_o''\neq N_i''}
\;\; \Big\}
\;,
\eea
where
$\rho^{1/2}_H\otimes\rho^{1/2}_H$
denotes, respectively,
$\;\rho^{1/2}_{H}(\omega_{o^{\prime}},-\widetilde{\omega}_{ i^{\prime}})\otimes\rho^{1/2}_{H}(\omega_{o^{\prime\prime}},
-\widetilde{\omega}_{i^{\prime\prime}})\;$,
$\;\rho^{1/2}_{H}(\omega_{o^{\prime}},-\widetilde{\omega}_{i^{\prime\prime}})
\otimes\rho^{1/2}_{H}(\omega_{o^{\prime\prime}},
-\wt{\omega}_{ i^{\prime}})\;$  and  $\;\Delta\rho_{{OD}}\otimes\left(\Delta\rho_{{OD}}\right)^\dagger$ denotes, respectively,
$\;\frac{1}{2}\left[\Delta\rho_{{OD}} (\omega_{o^{\prime}},-\widetilde{\omega}_{i^{\prime}}) \otimes
\Delta\rho_{{OD}}^{\dagger}(\omega_{o^{\prime\prime}},
-\widetilde{\omega}_{i^{\prime\prime}})
+ \Delta\rho_{{OD}}(\omega_{o^{\prime\prime}},
-\widetilde{\omega}_{i^{\prime\prime}}) \otimes
\Delta\rho_{{OD}}^{\dagger} (\omega_{o^{\prime}},-\widetilde{\omega}_{i^{\prime}})\right]
\;$,
$\;\frac{1}{2}\left[\Delta\rho_{{OD}} (\omega_{o^{\prime}},-\widetilde{\omega}_{i^{\prime\prime}}) \otimes\Delta\rho_{{OD}}^\dagger(\omega_{o^{\prime\prime}},
-\widetilde{\omega}_{i^{\prime}})
+ \Delta\rho_{{OD}}(\omega_{o^{\prime\prime}},
-\widetilde{\omega}_{i^{\prime}})\otimes \Delta\rho_{{OD}}^{\dagger} (\omega_{o^{\prime}},-\widetilde{\omega}_{i^{\prime\prime}})\right]
\;$.
The expression $\frac{1}{2}\left[\Delta\rho_{{OD}}\otimes \rho^{1/2}_H
+\rho^{1/2}_H \otimes
\left(\Delta\rho_{{OD}}\right)^\dagger\right]$ denotes, respectively,
the following:
$\;\frac{1}{2} \left[\Delta\rho_{{OD}}(\omega_{o^{\prime}},-\widetilde{\omega}_{i^{\prime}}) \otimes
\rho^{1/2}_H(\omega_{o^{\prime\prime}},
-\widetilde{\omega}_{i^{\prime\prime}})
+ \rho^{1/2}_H(\omega_{o^{\prime\prime}},
-\widetilde{\omega}_{i^{\prime\prime}}) \otimes
\Delta\rho_{{OD}}^{\dagger} (\omega_{o^{\prime}},-\widetilde{\omega}_{i^{\prime}})\right]
\;,\;\;\;$
$\;\frac{1}{2} \left[\Delta\rho_{{OD}}(\omega_{o^{\prime}},-\widetilde{\omega}_{i^{\prime\prime}}) \otimes\rho^{1/2}_H(\omega_{o^{\prime\prime}},
-\widetilde{\omega}_{i^{\prime}})
+ \rho^{1/2}_H(\omega_{o^{\prime\prime}},
-\widetilde{\omega}_{i^{\prime}})\otimes \Delta\rho_{{OD}}^{\dagger} (\omega_{o^{\prime}},-\widetilde{\omega}_{i^{\prime\prime}})\right]
\;,\;\;\;$
$\;\frac{1}{2} \left[\Delta\rho_{{OD}}(\omega_{o^{\prime\prime}},-\widetilde{\omega}_{i^{\prime}}) \otimes\rho^{1/2}_H(\omega_{o^{\prime}},
-\widetilde{\omega}_{i^{\prime\prime}})
+ \rho^{1/2}_H(\omega_{o^{\prime}},
-\widetilde{\omega}_{i^{\prime\prime}})\otimes \Delta\rho_{{OD}}^{\dagger} (\omega_{o^{\prime\prime}},-\widetilde{\omega}_{i^{\prime}})\right]
\;,\;\;\;$
$\;\frac{1}{2} \left[\Delta\rho_{{OD}} (\omega_{o^{\prime\prime}},-\widetilde{\omega}_{i^{\prime\prime}}) \otimes\rho^{1/2}_H(\omega_{o^{\prime}},
-\widetilde{\omega}_{i^{\prime}})
+ \rho^{1/2}_H(\omega_{o^{\prime}},
-\widetilde{\omega}_{i^{\prime}})\otimes \Delta\rho_{{OD}}^{\dagger} (\omega_{o^{\prime\prime}},-\widetilde{\omega}_{i^{\prime\prime}})\right]
\;$.

The out-particle reduced density matrix is obtained by tracing over the frequencies and particle numbers of the in-mode Hilbert space,
$\;\rho_{SC,out}={\rm Tr}_{in}\; \rho_{SC,o\otimes i}\;$. This is
a  straightforward calculation for the Hawking
part of the matrix. For the correction, it entails computing the
integral
\be
{\cal I}_{b}\; = \;
 \int\limits^{N_T}_{N_{T}-N_{coh}(N_T)} dN_i\;
e^{- \frac{1}{4} b\frac{(N_T-N_i)^2}{ N^2_{coh}(N_T;N_i)}}\;,
\label{calI}
\ee
where  $\;N_{coh}(N_T;N_i)\gg 1\;$ allows us to treat the discrete sum as continuous and
 $b$ is either 0, 1 or 2. For the terms
of order $C_{BH}^{1/2}$, then  $b$ is 0 or 1
and, for the terms of order $C_{BH}$,
then $b$ depends    on which of the four different
products of exponents  is being considered in the
product of  suppression factors,
\be
\frac{1}{4}\Big[ e^{-\frac{1}{4} \frac{(N_T-N_o^\prime)^2}{ N^2_{coh}(N_T;N_o^{\prime})}}+
e^{-\frac{1}{4}\frac{ (N_T-N_i)^2}{ N^2_{coh}(N_T;N_i)}}\Big]\times\Big[
e^{-\frac{1}{4} \frac
{(N_T-N_o^{\prime\prime})^2}{ N^2_{coh}(N_T;N^{\prime\prime}_o)}}+ e^{-\frac{1}{4}\frac{(N_T-N_i)^2}{N^2_{coh}(N_T;N_i)}}\Big]\;.
\ee

We will consider, for the most part, ``typical'' times $t$ such that $\;t_{coh}< t < \tau_{BH}-t_{coh}\;$. Then
the dependence on the second argument in $\;N_{coh}(N_T;N_i)\;$ is weak, $\;N_{coh}(N_T;N_i)\simeq N_{coh}(N_T)\equiv N_{coh}(N_T;N_T)\;$ for the relevant values of $N_i$. This
is made clear in the next paragraph.

It is obvious that  $\;{\cal I}_{b=0}=N_{coh}(N_T)\;$. For $\;b=1,2\;$, we note that $C_{BH}(N_T)$ is approximately constant over one interval of
coherence time. This is because $\;C_{BH}(N_T)\simeq \left[S_{BH}(0)-N_T\right]^{-1}\;$ \cite{slowleak}, from which it follows that
the change of $C_{BH}$ over a coherence time is small,
$\;\Delta C_{BH}\simeq
\frac{\partial C_{BH}}{\partial N_T}N_{coh}\simeq C_{BH}^2 N_{coh}\ll 1\;$.
Consequently, the integrand in ${\cal I}_{b=1,2}$ is approximately unity
over the range of integration.
It can be concluded that  $\;{\cal I}_{b}\simeq N_{coh}(N_T)\;$ for $\;b=0,1,2\;$.

It is now  straightforward to evaluate the reduced
density matrix,
\bea
\rho_{SC,\;out}(N_T;N_o',N_o'') &=&\frac{2N_{coh}(N_T)}{n^2_*}
{\rm Tr_{in}}[\rho_H] |
N_o'\rangle \langle N_o''|
\delta_{N_o',N_o''} \label{rhoscout}
 \\
&+& C_{BH}(N_T)\frac{N_{coh}(N_T)}{2n^2_*}  {\rm Tr_{in}}[\left|\Delta\rho_{{OD}}\right|^2]
\;\wt D(N_T;N_o',N_o'')|
N_o'\rangle \langle N_o''| \nonumber \\
&+& C^{1/2}_{BH}(N_T)\frac{N_{coh}(N_T)}{n^2_*}  {\rm Tr_{in}}[\overline{\Delta\rho}_{{OD}}]
\;\overline{D}(N_T;N_o',N_o'')
|N_o'\rangle \langle N_o''|_{N_o'\neq N_o''}
\;,
\nonumber
\eea
where
\be
\wt D(N_T;N',N'')\;\equiv\;\left[1+ e^{-\frac{1}{4} \frac{\left[C_{BH}(N^\prime) (N_T-N^\prime)\right]^2}{ C_{BH}(N_T)}}\right]\times
\left[1+ e^{-\frac{1}{4} \frac{\left[C_{BH}(N^{\prime\prime}) (N_T-N^{\prime\prime})\right]^2}{ C_{BH}(N_T)}}\right]\;,
\label{wtD}
\ee
\be
\overline{D}(N_T;N',N'')\;\equiv\;
e^{-\frac{1}{4} \frac{\left[C_{BH}(N')(N_T-N')\right]^2}{C_{BH}(N_T)}}
+2+e^{-\frac{1}{4} \frac{\left[C_{BH}(N'') (N_T-N'')\right]^2}{ C_{BH}(N_T)}}\;,
\ee
\be
\overline{\Delta\rho}_{{OD}}\;=\;\frac{1}{2}\left[\Delta\rho_{{OD}}\otimes \rho_H^{1/2} +\rho_H^{1/2}\otimes \left(\Delta\rho_{{OD}}\right)^{\dagger}\right]\;
\ee
and
$\;{\rm Tr_{in}}\left[\rho(\omega_{o'},-\wt{\omega}_{ i'})\rho^\dagger(\omega_{o''},-\wt{\omega}_{i''})
\right]
=\int\limits^{\infty}_{0} dx\; \rho(\omega_{o'},-x)\rho^\dagger(\omega_{o''},-x) \;$.
Here, $\rho$ means any single-pair matrix and
$\;{\rm Tr_{in}}\left[\rho^{1/2}_H\rho_H^{1/2}\right]={\rm Tr_{in}}[\rho_H]\;$
has been used.

To fix the normalization constant $n_*$ in Eq.~(\ref{rhoscout}),  we need to calculate the trace of $\rho_{SC,\;out}$. We will assume the convention that
the full  trace over the single-pair Hawking matrix gives unity,
$\;{\rm Tr_{out}Tr_{in}}[\rho_H]=1\;$, thus absorbing the implicit correction
of order $C_{BH}^{1/2}$ for the  diagonal ($N'=N''$) terms  into the
normalization.

The trace of the first term in $\rho_{SC,\ out}$ is given by
\be
\frac{2N_{coh}(N_T)}{n^2_*}
{\rm Tr_{out}}\left[ {\rm Tr_{in}}[\rho_H] |
N_o'\rangle \langle N_o''|
\delta_{N_o',N_o''}\right]\;=\; \frac{2N^2_{coh}(N_T)}{n^2_*}\;.
\ee

For the calculation of the  trace of the  correction,
the relevant integral is
\be
{\cal J}\;=\;\int\limits^{N_T}_{N_T-N_{coh}(N_T)}\;dN\; \wt D(N_T;N;N)\;.
\ee
This integral has four contributions, each of which is of
the same form as ${\cal I}_{b}$ in Eq.~(\ref{calI}).
Then, by the  same reasoning, $\;{\cal J}\simeq 4N_{coh}\;$.  Hence,
\be
\hspace*{-.5in} C_{BH}(N_T)\frac{N_{coh}(N_T)}{2n^2_*}
{\rm Tr_{out}}\left[   {\rm Tr_{in}}[\Delta\rho^2_{{OD}}]
\wt D(N_T;N_o',N_o'')|
N_o'\rangle \langle N_o''|\right]\;=\;\gamma  C_{BH}(N_T) \frac{2N^2_{coh}(N_T)}{n^2_*}\;,
\ee
whereby
\be
\gamma\;\equiv\;{\rm Tr_{out}Tr_{in}}[\left|\Delta\rho_{{OD}}\right|^2]\;,
\ee
with $\gamma$ being a number of order unity.

Then, as the full trace of the density matrix should be unity, it follows that
\be
\;n^2_*=2N^2_{coh}(N_T)\left[1+\gamma\;C_{BH}(N_T)\right]\;.
\ee

Our starting point for the next section is the reduced density matrix for the out-modes with
correct normalization (up to order $C_{BH}$),
\bea
\rho_{SC,\;out}(N_T;N_o',N_o'') &=&\frac{1}{N_{coh}(N_T)}\left[1- \gamma\;
C_{BH}(N_T)\right]
{\rm Tr_{in}}[\rho_H] |
N_o'\rangle \langle N_o''|
\delta_{N_o',N_o''} \nonumber
\\
&+& \frac{C_{BH}(N_T)}{4N_{coh}(N_T)}  {\rm Tr_{in}}[\left|\Delta\rho_{{OD}}\right|^2]
\;\widetilde{D}(N_T;N_o',N_o'')|
N_o'\rangle \langle N_o''|
 \nonumber \\
&+& \frac{C^{1/2}_{BH}(N_T)}{2N_{coh}(N_T)}  {\rm Tr_{in}}[\overline{\Delta\rho}_{{OD}}]
\;\overline{D}(N_T;N_o',N_o'')|
N_o'\rangle \langle N_o''|_{N_o'\neq N_o''}
 \nonumber
\\ && \label{redmat}
\eea

\section{In-Out entanglement}

To determine the entanglement entropy for an approximately
pure state, it is appropriate to use the von Neumann entropy formula.
Then the entanglement entropy per particle is~\footnote{The von Neumann
formula only gives the total entanglement if the particles had
first been symmetrized.}
\be
\frac{S_{ent}}{N_{part}}= -{\rm Tr}[\wh{\rho}\ln{\wh{\rho}}]\;,
\ee
where, in our case, $\;N_{part}=N_{coh}\;$.

For the Hawking part alone,
\be
[S_{ent}]_H \;=\;-\left[1-\gamma\;C_{BH}(N_T) \right]
{\rm Tr}_{out}{\rm Tr}_{in}\left[\rho_H\ln{\rho_H}\right]\;
N_{coh}(N_T)\;,
\ee
where  we have eliminated
a factor of  $\ln(N_{coh})$  by  correcting  for Gibbs' paradox for
indistinguishable particles.  The leading-order outcome  is, of course,
the expected
result for a total of $N_{coh}$ maximally entangled pairs.

\subsection{In-Out entanglement for $t<t_{trans}$}

What is left to resolve is the effect of the correction.
Let us  recall that we are considering a ``typical'' BH for $t_{coh}<t<t_{trans}$. The extreme cases will be addressed further along.

For a matrix of the form of that in  Eq.~(\ref{redmat}), the effective
perturbation parameter is $\;C_{BH}N_{coh}\sim C^{1/2}_{BH}\ll 1\;$ \cite{slowburn}. This  is because the corrections of  order $C_{BH}^{1/2}$ are
strictly off-diagonal and so can only appear at quadratic order in tracing operations. Hence, we can evaluate the correction to the entanglement perturbatively.

Then, to  proceed, we  expand
out the logarithm in the von Neumann formula to linear  order in $C_{BH}$  and
use the approximation $\;{\cal J}\simeq 4N_{coh}\;$ (and its $\overline D$ analogue)
as discussed above. The result  (again after accounting for Gibbs'
paradox)
is
\be
S_{ent} \;=\; -{\rm Tr}_{out}{\rm Tr}_{in}\left[\rho_H\ln{\rho_H}\right]N_{coh}(N_T)\Big(1\;-\;\frac{(\gamma+1)\;C_{BH}(N_T)}{-{\rm Tr}_{out}{\rm Tr}_{in}\left[\rho_H\ln{\rho_H}\right]}\Big)\;.
\ee
One can observe that the entanglement of the in--out sector is still
parametrically close to  maximal,  $\;S_{ent}(N_T)\sim N_{coh}(N_T)\;$.

\subsection{Qualitative analysis of the in--out entanglement for $t>t_{trans}$}

Let us now remove the constraint of typicality on the age of the
BH. Early times in the evolution ($t<t_{coh}$) are well understood and need
not concern us, but what about late in the process?
From the analysis in \cite{slowburn}, we have  observed that the
BH evolves in typical fashion until about one interval of coherence time before the end of evaporation. At this point, which is what we  call the transparency time $\;t_{trans}$, $C_{BH}$ starts growing rapidly from a value of  $\;C_{BH}(t_{trans})\sim S^{-2/3}_{BH}(0)\;$ to its value of unity for a Planck-sized
BH.  Moreover, the information  $I$ begins to rapidly emerge from the BH, $\;\left.\frac{dI}{dN_T}\right|_{t_{trans}}\sim 1\;$.

Since $C_{BH}$ is becoming large at such late times, it is evident that our previous perturbative treatment is no longer applicable. In fact, as made clear in \cite{slowburn}, our treatments already begins to  break down at the transparency time. However, by this time, at least  $\;\left[N_T\right]_{max}- N_{coh}(t_{trans})\simeq S_{BH}(0)-N_{coh}(t_{trans})\;$ particles have already been radiated away. So the number of Hawking particles which are yet to be emitted is about $\;N_{coh}(t_{trans})\simeq S^{1/2}_{BH}(t_{trans})\sim S_{BH}^{1/3}(0)\;$. This means that, from $t_{trans}$ until the BH evaporates completely, at most $S_{BH}^{1/3}(0)$ entangled pairs can be created. It is quite possible that these remaining pairs are no longer maximally entangled. However, the BH is now within the final stage of evaporation and  there is no compelling reason to believe that the late-time  horizon is cold. Any chance of forming a firewall is delayed at least until a parametrically smal
 l time before the BH has finally evaporated.

Let us now consider in more detailed way the evolution of the entanglement  for $t>t_{trans}$. As already remarked upon, we cannot make precise statements as to what transpires at times later than $t_{trans}$. In particular (and as discussed in \cite{slowburn}), we do not know the precise expression for the coherence scale $N_{coh}(N_T;N')$ in this regime.

On the other hand, contrary to our previous out--out analysis,  the precise value of
the coherence scale is not particularly relevant to the in--out sector at late times. This is because the number of remaining coherent  pairs $N_{pairs}$
 cannot be larger than the total number of Hawking particles remaining to be emitted, and so $N_{pairs}$
is no longer set
by the coherence scale but rather   $\;N_{pairs}(N_T)=[N_T]_{max}-N_T\;$.
 We find this number to be smaller than  the coherence scale, $\;N_{pairs}(N_T) < N_{coh}(N_T;N')\;$, for $\;t > t_{trans}\;$. This follows from the qualitative estimates in \cite{slowburn}, where $N_{coh}(N_T;N')$ was found to be  a  monotonically growing quantity after the transparency time. The only exception
being  $\;N' \simeq N_T\simeq N_{trans} \;$; in which case,  $N_{coh}$ and $N_{pairs}$ are parametrically similar.

During these late times, $C_{BH}(N_T)$ is becoming large and will eventually approach unity as the BH tends toward Planckian dimensions. As a consequence, any of the exponential  suppression factors ({\em e.g.}, the first one in  Eq.~(\ref{wtD})) becomes like a  theta  or Heaviside  function, as the  numerator of the exponent $\left[C_{BH}(N')(N_T-N')\right]^2$ is a number of order unity for all choices  of  $N'$ when  $N_T$ is approaching its maximum value.

In light of the above, the reduced density matrix for
the out-modes simplifies at late times,
\bea
\rho_{SC,\;out}(N_T;N_o',N_o'') &=&\frac{1}{N_{pairs}(N_T)}
{\rm Tr_{in}}[\rho_H] |
N_o'\rangle \langle N_o''|
\delta_{N_o',N_o''} \nonumber
\\
&+& \frac{2}{N_{pairs}(N_T)} C^{1/2}_{BH}(N_T)
 {\rm Tr_{in}}[\overline{\Delta\rho}_{{OD}}]
|N_o'\rangle \langle N_o''|_{N''_o\neq N'_o}\nonumber
\;  \\
&+& {\cal O}[C_{BH}]\;,
\eea
with $C_{BH}$  now  regarded as a number
that is still small (relative to unity) but large enough
to satisfy  $\;C_{BH}\gg N^{-1}_{coh}\;$.

The ``correction'' part of the matrix seems to become more important and eventually would seem dominate the diagonal part. This is because the Hawking part is diagonal with  $N_{pairs}$ entries, whereas the correction is a nearly uniform matrix  with   $\;N_{pairs}^2-N_{pairs}\gg N_{pairs}\;$ entries. However, notice  that the off-diagonal part can only appear quadratically when a trace of some operator is evaluated. And so the correction it makes to the entropy or other physical quantities is actually suppressed by a power of $\;C_{BH} <1\;$ with respect to the diagonal contribution until such time as the BH approaches Planckian dimensions. Hence, we can expect the previous (early-time) calculation  to remain roughly valid at least until the BH has shrunk  past its regime of semiclassical validity. We can then conclude that, even at late times,  $\;S_{ent}\sim N_{pairs}\;$ and the in--out  entanglement  remains parametrically close to maximal, while monotonically decreasing
  in time in the same way that  $N_{pairs}$ does.

As already emphasized, these arguments are qualitative.  It
is, however, worthwhile to remember that the late-time horizon region
can not necessarily be expected to be similar to the vacuum, contrary
to our expectations at earlier times. Hence, our overall argument
does not hinge on the exact fate of  the in--out sector at these
final stages.

\subsection{Summary of the evolution of entanglement entropy and released information}

Let us summarize the dependence of the in--out entanglement on time.
For $t<t_{coh}$ the entanglement entropy increases linearly with the number of emitted Hawking particles. For $t_{coh}<t<t_{trans}$, the entanglement entropy is equal to $N_{coh}$,  which is a very slowly decreasing function of the number of emitted particles and  can then be approximated by a constant. After $t_{trans}$,  the entanglement entropy decreases to zero. We have argued that this decrease is linear in the number of particles that are yet to be emitted.

The following equation summarizes the different dependencies of the entanglement entropy:
\begin{equation}
S_{ent}(N_T)\sim\begin{cases}
N_T & 0\le N_T \le N_{coh} \cr
N_{coh}\sim \sqrt{N-N_T} & N_{coh}\le N_T \le N-N^{2/3}  \cr
N-N_T & N -N^{2/3}\le N_T\le N \;.
\end{cases}
\end{equation}
Here, we have denoted by $N$ the total number
of Hawking particles emitted during the lifetime of the BH, $\;N=\left[N_T\right]_{max}\simeq S_{BH}(0)\;$. (This $N$ should not to be confused with the argument of $N_{coh}$.)

For comparison, we  also recall  how $S_{ent}$ evolves  for the Page model
\cite{page},
\begin{equation}
S_{ent}^{Page}(N_T)\sim\begin{cases}
N_T & 0\le N_T \le N/2
\cr N-N_T & N/2 \le N_T\le N\;.
\end{cases}
\end{equation}

\begin{figure}[ht]
\ \hspace{1in}\scalebox{.99} {\includegraphics{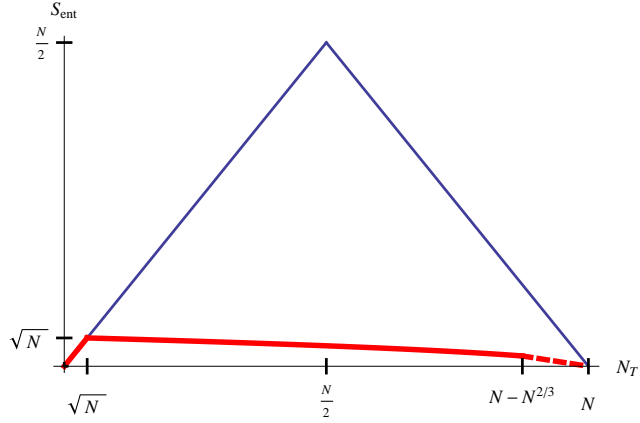}}
\caption{\small Entanglement entropy $S_{ent}$ of the Hawking pairs, as a function of the number  of emitted Hawking particles $N_T$.  This is shown both for the Page model (thin, blue) and ours
(thick, red). The units are arbitrary. The main
difference is the maximal value of the entanglement entropy, which scales as $N_T$ for the Page model and as $\sqrt{N_T}$ for ours. The decrease in $S_{ent}$ starts when $dI/dN_T\sim 1$.}
\end{figure}

Let us further recall how the released information $I$
depends on the number of emitted particles, as calculated in \cite{slowburn},
\begin{equation}
I(N_T)\;\sim
\begin{cases}
\;N_T \frac{1}{(N-N_T)^{1/2}} & N_T \ll N-N^{2/3}
\cr \;N_T \frac{N}{(N-N_T)^{3/2}} & N_T \lesssim N-N^{2/3}
\end{cases}
\label{INFO}
\end{equation}
and
\be
\frac{dI}{dN_T} \;\sim\; 1 \hspace{.50in} N_T \sim N-N^{2/3}\;.
\label{INFOR}
\ee

For comparison, the evolution of the released information  for the  Page model
goes as
\begin{equation}
I^{Page}(N_T)\sim\begin{cases} 0 & 0\le N_T \le N/2 \cr 2(N_T-N/2) & N/2 \le N_T\le N
\;. \end{cases}
\end{equation}
The dependence of $S_{ent}$ on the number of emitted particles is shown in
Fig.~1.

It should be emphasized that we are only considering the entanglement between Hawking modes and their negative-energy partners and not the entanglement between the Hawking modes and the rest of the interior of the BH. This is the reason that the graph in Fig.~1 takes the flattened form that it does. This distinction between partners and the interior of the BH is of no consequence to Page because of his indifference to the horizon. It does, however, make a difference for us because our framework is such that it limits the number of negative-energy partners at any given time to $\;N_{coh}\sim S^{1/2}_{BH}\;$;  which necessarily limits the amount of entanglement in the same way. Physically, the semiclassical horizon is acting to shield all but a fraction $N_{coh}$ of the matter modes.

On the other hand, the breakdown of the semiclassical picture at late times
(see Subsection~1.1 for a discussion)
suggests that all the information can still be released once the
(quantum) horizon is
no longer
acting  as a causal barrier. Some qualtitative estimates in \cite{slowburn} along with
Eq.~(\ref{INFOR}) are in  support of this argument. We do, however, expect to put this claim on a more
rigorous level at a later time \cite{endgame}.

The most significant  physical difference between our model and the Page model is the incorporation (or not) of a horizon. Page basically disregards the presence of a horizon and treats the BH evaporation in purely information-theoretic terms. He assumes that the off-diagonal elements of the density matrix are distributed randomly, with strength controlled by the dimensionality of the entire BH Hilbert space. On the other hand, our framework is based on the presence of a semiclassical horizon. Had we treated the horizon classically, there would be no off-diagonal elements as is the case in Hawking's calculation. But, because our horizon is semiclassical, not all the off-diagonal elements are vanishing. The number of non-vanishing off-diagonal elements is controlled by the wavefunction of the BH; specifically, the width of the Gaussian.  The density matrix and its physical consequences then follows. This leads us to a density matrix with many
zeroes,  which would be viewed by Page as a very atypical matrix.

Because Page essentially ignores the horizon,
the distinction between partners and interior matter  is of no consequence to
his model. It does, however, make
a difference for us because our framework is such that it limits the
number of negative-energy partners at any given time  to $\;N_{coh}\sim S^{1/2}_{BH}\;$, which necessarily limits the amount of entanglement in the same way.
This upper bound on the entanglement explains  the flattening of the
red curve in Fig.~1.
Physically, the semiclassical horizon  is acting to shield all but a fraction
$N_{coh}$ of
the matter modes. This being  a consequence of  our choice of
BH wavefunction, which determines  the
transparency of the  horizon.

\section{Discussion}

We have  shown that the in--out sector of the BH radiation  is close to maximally entangled; at least  until the transparency time, when our perturbative analysis begins to break down. Additionally,  the entanglement between in- and out-modes is limited to a maximal value of  $\;N_{coh}\simeq \sqrt{S_{BH}}\;$ that is parametrically smaller than the total number of emitted Hawking particles. This limitation can be attributed to the regular recycling of partnered modes over a time scale that is set by the quantum width of the wavefunction for the evaporating BH.

The limited dimensionality of the boundary region between the interior and exterior Hilbert spaces is central. This  restriction can be attributed to incorporating  both the wavefunction for the BH geometry and the  back-reaction on the BH due to the emitted particles. When $\;t<t_{trans}\;$, the semiclassical corrections to the in--out  density matrix are insignificant --- the Hawking pairs are highly entangled with or without them. On the other hand,  these corrections enable information to  be transferred to  the  outgoing radiation via the off-diagonal corrections to the Hawking matrix. Their presence follows from treating the BH consistently as a quantum object.

Let us recall \cite{slowburn}, where qualitative considerations were used to conclude  that the outgoing radiation starts to purify at the same late time when the entanglement entropy is beginning to decrease. The implication for  our framework is that, for times earlier than  the transparency time, any  duplication of entanglement or purity \cite{bousso2} is avoided without the need to modify the rules of quantum mechanics. Hence, our  framework is immune against the formation of firewalls and horizons are cold, at least until one coherence time before the end of evaporation.

Let us reconsider the Page model and its associated firewall. The implicit assumption in this model is that the number of strongly entangled pairs becomes of order  $S_{BH}$ by the Page time \cite{page}. But this is also supposed to be the time  when information begins to rapidly emerge from the BH, and one encounters the inevitable conflict of interests. The transfer of information  means the transfer of entanglement from the partnered pairs to the late--early radiation. One is then faced with the prospect of a firewall or, otherwise, a means for circumventing the rule about monogamy of entanglement. As just mentioned, in our model of BH evaporation, this issue is postponed until a much later time, when the BH stops being semiclassical.
In our model, the interior system that purifies the outgoing radiation has an additional component, the BH wavefunction. This additional component is likely representing the collapsed matter from an in-falling observer's perspective.

Furthermore, even for our model, the in--out entanglement  does stray from its maximal value  by small amounts. This result suggests the interesting possibility that some part of the firewall idea still survives.  Very old BHs do seem to have different properties than younger ones. The question then arises: How much of a deviation is needed before a firewall forms?  A related question is how strongly did our conclusions depend on the precise choice for the wavefunction and on our model of back-reaction. We hope to make these questions more precise and provide quantitative answers in a future article \cite{noburn}.

\section*{Acknowledgments}

We thank  Sunny Itzhaki for discussions and useful comments on the manuscript. The research of RB was supported by the Israel Science Foundation grant no. 239/10.
The research of AJMM received support from an NRF Incentive Funding Grant 85353 and
an NRF KIC Grant 83407.
AJMM thanks Ben Gurion University
for their hospitality during his visit.

\end{document}